\documentclass{article}
\usepackage{spconf,amsmath,graphicx,hyperref}

\usepackage{url}
\usepackage{graphicx}
\usepackage{cite}
\usepackage{amsmath,amssymb,amsfonts}
\usepackage{algorithmic}
\usepackage{graphicx}
\usepackage{textcomp}
\usepackage{xcolor}
\definecolor{dkgreen}{rgb}{0,0.6,0}
\definecolor{gray}{rgb}{0.5,0.5,0.5}
\definecolor{mauve}{rgb}{0.58,0,0.82}
\graphicspath{{Figure/}}
\usepackage{cite}
\usepackage{amsmath,amssymb,amsfonts}
\usepackage{algorithmic}
\usepackage{graphicx}
\usepackage{textcomp}
\usepackage{xcolor}
\usepackage{booktabs}
\usepackage{multirow}
\usepackage{lscape}
\usepackage{graphicx}
\usepackage{longtable}
\usepackage{supertabular,booktabs}
\usepackage{subfig}
\usepackage{subcaption}
\usepackage{comment}
\usepackage[normalem]{ulem}
\useunder{\uline}{\ul}{}
\usepackage{colortbl}
\usepackage{graphicx}
\usepackage{hyperref}

\title{DAT-CFTNet: Speech Enhancement for Cochlear Implant Recipients using Attention-based Dual-Path Recurrent Neural Network}
\name{Nursadul Mamun$^{1}$, John H.L. Hansen$^{2}$}
\address{
$^{1}$Chittagong University of Engineering and Technology, Chittagong, Bangladesh \\ $^{1,2}$CRSS: Center for Robust Speech Systems, University of Texas at Dallas, USA \\
\texttt{nursad.mamun@cuet.ac.bd, john.hansen@utdallas.edu}
}
\begin{document}
%
\maketitle

\begin{abstract}
The human auditory system has the ability to selectively focus on key speech elements in an audio stream while giving secondary attention to less relevant areas such as noise or distortion within the background, dynamically adjusting its attention over time. Inspired by the recent success of attention models, this study introduces a dual-path attention module in the bottleneck layer of a concurrent speech enhancement network. Our study proposes an attention-based dual-path RNN (DAT-RNN), which, when combined with the modified complex-valued frequency transformation network (CFTNet), forms the DAT-CFTNet. This attention mechanism allows for precise differentiation between speech and noise in time-frequency (T-F) regions of spectrograms, optimizing both local and global context information processing in the CFTNet. Our experiments suggest that the DAT-CFTNet leads to consistently improved performance over the existing models, including CFTNet and DCCRN, in terms of speech intelligibility and quality. Moreover, the proposed model exhibits superior performance in enhancing speech intelligibility for cochlear implant (CI) recipients, who are known to have severely limited T-F hearing restoration (e.g., \(>\)10\%) in CI listener studies in noisy settings show the proposed solution is capable of suppressing non-stationary noise, avoiding the musical artifacts often seen in traditional speech enhancement methods. The implementation of the proposed model will be publicly available\footnote{\url{https://github.com/nursad49/DAT-CFTNet}}
\end{abstract}

\section{Introduction}
\label{sec:intro}

Cochlear implants (CI) provide a valuable solution for individuals with severe hearing loss, allowing them to experience sound by directly stimulating the auditory nerve \cite{zeng2008cochlear}. However, CI users often face challenges in noisy environments where speech can be masked with widespread background noise \cite{mamun2019convolutional}. This limitation can reduce the overall quality of life and hinder effective communication for CI recipients. Traditional speech enhancement (SE) techniques aim to extract the clean speech signal from an input noisy observation, thereby reducing the effects of the background noise. The primary objective of SE for CIs is not only noise reduction but also the preservation and amplification of speech intelligibility and quality given that CI subjects only receive $\sim$10\% of the T-F content vs. that of normal hearing (NH) subjects. Given the distinct auditory processing of CI recipients, traditional SE algorithms will certainly not yield optimal results. Therefore, there is a pressing demand to formulate and evaluate techniques designed specifically for CI users tailored to their unique reduced 10\% T-F hearing capabilities.


In recent decades, deep-learning-driven single-channel SE methods have achieved remarkable success, especially in low SNR settings. While convolutional neural networks (CNNs) \cite{park2016fully, mamun2024speech, mamun2021self, mamun2019quantifying} excel in representation, they struggle with long-range dependencies. Recurrent neural networks (RNNs) \cite{valentini2016investigating}, including LSTMs, effectively model long-term sequences but are computationally intensive due to their sequential processing. To harness the strengths of both, hybrid models such as convolutional recurrent neural networks (CRN), deep complex convolution recurrent neural (DCCRN) networks \cite{hu2020dccrn}, and gated convolution recurrent neural (GCRN) networks \cite{tan2019learning} have been developed. These networks aim to capture both local and long-range information. However, they still face challenges in restoring certain speech frequency components, impacting the final speech-to-distortion ratio.


To address this, a complex frequency transformation network (CFTNet) has recently been proposed in \cite{mamun2023sCFTNet}. Taking advantage of both UNet and frequency transformation layers, CFTNet captures global correlations over frequency for time-frequency (T-F) representations and allows the network to use limited frequency information to reconstruct missing frequency components in the distorted signals. However, the traditional RNN units embedded in its bottleneck layer fall short of effectively modeling extended speech feature sequences.

The dual-path recurrent neural network (DPRNN) adeptly manages long sequential inputs, especially in the T-F spectrum by synergizing the capabilities of intra-chunk and inter-chunk RNNs \cite{luo2020dual}. Inspired by the efficacy of DPRNN as a bottleneck layer in the dual-path convolutional recurrent network (DRCRN) \cite{le2021dpcrn} and the transformative impact of attention mechanisms in SE networks \cite{hao2019attention}, we present the dual-path attention RNN (DAT-RNN). Further, we seamlessly integrate DAT-RNN into CFTNet, leading to the creation of DAT-CFTNet. This innovation enhances the CFTNet by integrating an attention mechanism, leading to optimized memory usage and setting new benchmarks in SE performance. Unlike the CFTNet SE network \cite{mamun2023sCFTNet}, the DAT-CFTNet incorporates the DAT-RNN in its bottleneck layer. This integration leverages the benefits of capturing long-term dependencies, allowing the network to adeptly recognize both spectral subtleties and temporal dynamics distinct to each segment.


This paper is organized as follows: Sec. 2 provides a detailed implementation of our proposed DAT-CFTNet. The experimental setup and evaluation results are described in Sec. 3, followed by the conclusion in Sec. 4.

\section{Methodology}

\subsection{Dual-Path Attention CFTNet}

\begin{figure}[t]
  \centering
  \includegraphics[width=0.5\textwidth]{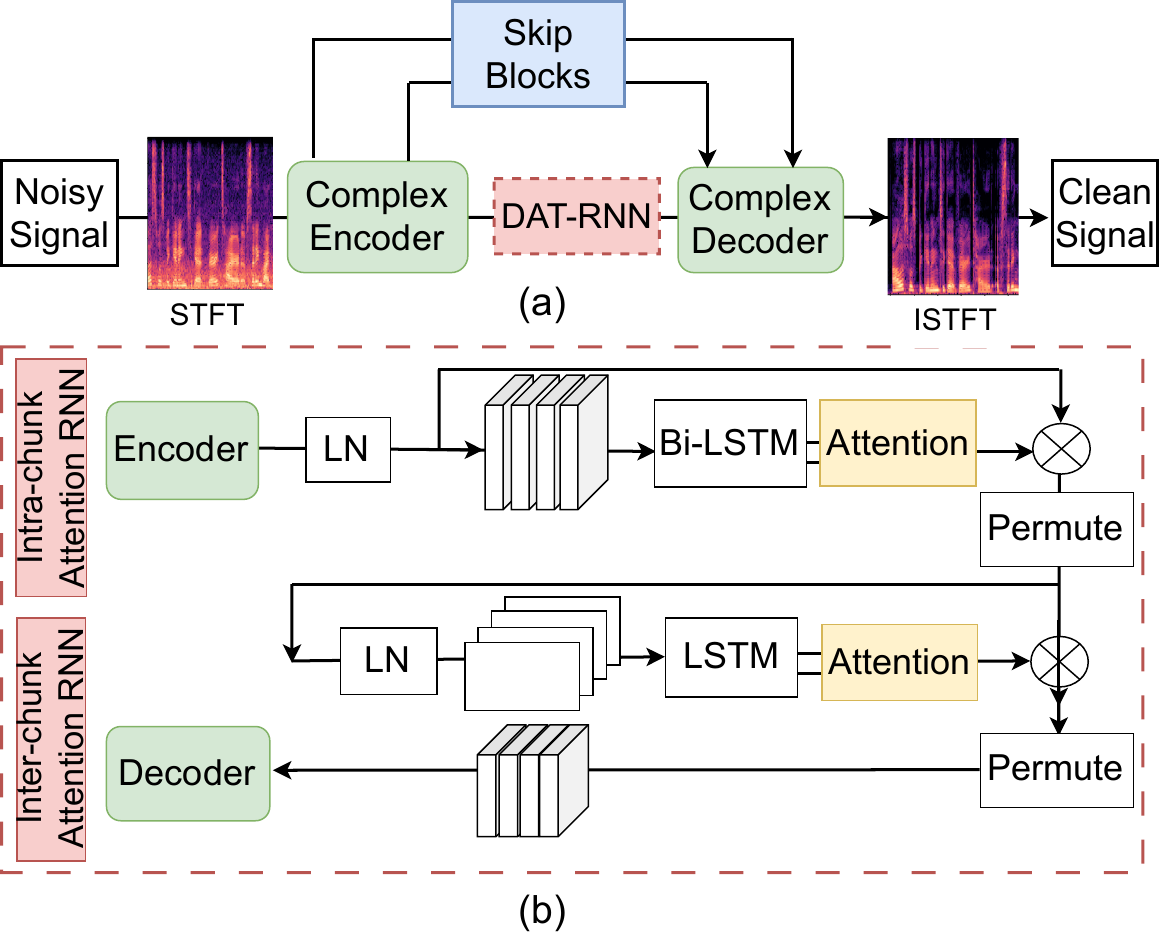}
  \caption{(a) Basic block diagram of the proposed DAT-CFTNet and (b) the dual-path attention RNN module}
  \label{fig: Network}
   \vspace{-20pt}
\end{figure}

 Figure~\ref{fig: Network} represents the block diagram of the proposed network. It comprises an encode, a decoder, and a DAT-RNN module in the bottleneck layer, mirroring the structure of CFTNet ~\cite{mamun2023sCFTNet}. The noisy spectrogram of is processed through complex-valued convolution layers for sequential enhancements in magnitude and phase. The Conv2D layers in the encoder extract local patterns and reduce feature dimensions, while its frequency transformation blocks (FTB) capture global correlations across T-F representations. The decoder, symmetrically designed to the encoder, reconstructs the clean spectrum from diminished features, with the skip block refining network learning.

Here, we have substituted the GRU module in CFTNet with the proposed DAT-RNN layer, as depicted in Fig. ~\ref{fig: Network}(b). The DAT-RNN module, merging DPRNN with an attention mechanism, adeptly models spectral patterns in T-F representations. This module integrates DPRNN with attention, fine-tuning spectral patterns in T-F representations. Like DPRNN, DAT-CFTNet employs two RNN types, each followed by an attention module. Within the spectrum, the intra-chunk RNN focuses on individual T-F units or `chunks', processing the intricate details and patterns within each localized T-F segment. This allows the network to understand spectral characteristics and temporal dynamics specific to each segment. By assigning varying attention weights to different parts of each chunk, the network can prioritize and capture more salient local features and nuances within that specific segment. Conversely, the inter-chunk RNN comes into play after the intra-chunk processing, aggregating the information across all the T-F chunks. Its primary function is to capture the overarching relationships and dependencies across these chunks, ensuring a holistic understanding of the entire T-F spectrum.

\begin{table*}
  \centering
  \caption{Mean objective scores for different networks for three SNRs.}
 \includegraphics[width=0.8\linewidth]{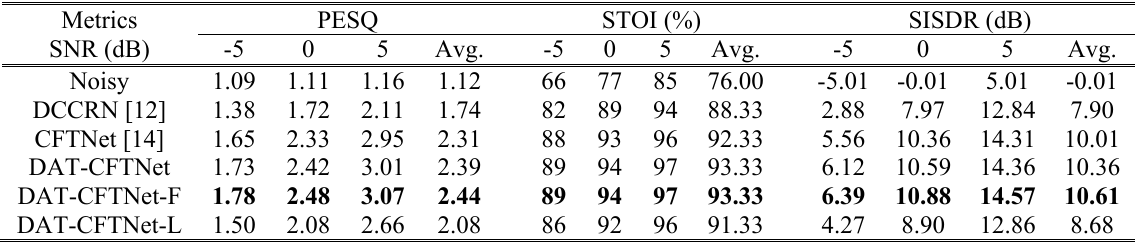}
 \label{table: Objective_Score}
 \vspace{-10 pt}
 \end{table*}
\subsection{Dual-Path Attention Module}
Next, Figure ~\ref{fig: Network}(b) illustrates the block diagram of the proposed DAT-RNN module. The DAT-RNN module is comprised of an intra-chunk RNN and an inter-chunk RNN module, with each having a subsequent attention module. Initially, the output from the encoder, denoted as X, is subjected to layer normalization (LN) to maintain a consistent input distribution. This normalized output is subsequently segmented into overlapping chunks. 

To extract intricate details from localized segments or chunks, the intra-chunk RNN employs bidirectional LSTM (Bi-LSTM). In contrast, the inter-chunk RNN, which uses standard LSTM, is designed to process information across all chunks, thereby capturing broader sequence patterns. Upon processing through the LSTM and Bi-LSTM layers, two crucial parameters, namely key and query, are forwarded to the attention module. This attention mechanism subsequently computes a mask vector, M, corresponding to the input feature. Utilizing this predicted mask vector, enhanced features corresponding to the original input features are then produced.


From the input, the LSTM and Bi-LSTM layers extract a high-level feature representation, denoted as $H_k^K$ and $H_k^Q$:
\begin{equation}
 \vspace{-10pt}
H_k^K, H_k^Q = layerNorm(f_e(X))
\end{equation} 
Here $f_e$ represents the LSTM or Bi-LSTM and $K$ and $Q$ represent the key and query, respectively \cite{vaswani2017attention}. A dynamic causal attention strategy calculates the normalized attention weight, W \cite{hao2019attention}. The attention mechanism processes the query and key to produce a context vector, $C_k$:
 \vspace{-10pt}
\begin{equation}
C_k = \sum_{k=1}^{t} {W_k H_k^K}
 \vspace{-15pt}
\end{equation}
\begin{equation}
W_k = \frac{f(exp(score(H_k^K, H_k^Q)))}{\sum_{k=1}^{t} {f(exp(score(H_k^K, H_k^Q)))}}
\end{equation}
where $f(.)$ represents the causal dynamic attention. The normalized attention mechanism is estimated. Then hidden mask vector is derived from the context vector $_k$ and the LSTM output $H_k^Q$, which is then used to estimate the enhanced features of each intra-chunk and inter-chunk RMM module.




 \vspace{-10pt}
\subsection{Depthwise Separable Convolution}
Depthwise separable convolution (DSC) is an efficient variant of standard convolution in CNNs, prominently used in lightweight models like MobileNets \cite{howard2017mobilenets}. It splits standard convolution into depthwise convolution (each input channel gets its filter) and pointwise convolution (1x1 convolution merging depthwise outputs). This approach significantly reduces computational needs and parameters, making it ideal for devices with limited resources. Its performance remains comparable to conventional convolution, especially in tailored architectures.
Our DAT-CFTNet enhances speech quality but has many trainable parameters, limiting real-time applications. To optimize this, we developed a DAT-CFTNet variant replacing standard Conv2D with DSC, leading to a threefold parameter reduction. 


 \vspace{-10pt}
\section{Experimental Setup}
\subsection{Speech Database}
\label{sec:database}

This study uses the IEEE database \cite{rothauser1969ieee}, with a original sampling frequency of 25 kHz and down-sampled to 16 kHz for this study. From this corpus, a subset of 1040 utterances from 104 sets was used for training. These sentences were augmented with nine distinct noise sources from the AURORA dataset \cite{hirsch2000aurora}, added at varying Signal-to-Noise ratios (SNRs) from -2 to 14 dB in 2 dB increments. We reserved 140 of these altered utterances from 14 sets as a validation test set. Environmental noise conditions ranged from large crowded noise and interior car noise type to speech-shaped noise (SSN) and white Gaussian noise. Furthermore, a second subset of 400 samples was tested under three seen (large crowd, car, SSN) and two unseen (restaurant, train) noise conditions at three SNR levels (-5, 0, and 5 dB).


 \vspace{-10pt}



\subsection{Network Architecture}
\label{sec:network}

The DAT-CFTNet is designed to estimate non-linear mappings between a noisy speech T-F spectrum and its clean counterpart. This process initiates with the computation of the speech signal's STFT, utilizing a frame size of 32 ms and a 16 ms overlap. The combination of CFTNet and DAT-RNN in the DAT-CFTNet model is denoted as DAT-CFTNet. Subsequently, we enhanced the CFTNet's encoder by strategically incorporating the FTB, leading to the birth of DAT-CFTNet-F. Instead of integrating FTB post each encoder block as in DAT-CFTNet, DAT-CFTNet-F positions two FTB after the first and final layers of the encoder block. This configuration assists the network in emphasizing the frequency components from the distorted T-F representation. A streamlined version of DAT-CFTNet-L adopts DSC over the traditional convolution, which trims the model parameters from 12.4 M to a mere 4.7 M.

\section{Results and Discussions}

This section presents an assessment of the performance of the proposed DAT-CFTNet, emphasizing objective metrics. The performance of DAT-CFTNet is evaluated using several measures, encompassing speech intelligibility, speech quality, and a speech distortion index. Subsequently, we compare these scores with those derived from well-established networks, including DCCRN and CFTNet, under various seen and unseen noise scenarios and SNRs. Additionally, we delve into the impact of distinct components within the proposed algorithm.

Table \ref{table: Objective_Score} illustrates the objective scores of the proposed network in terms of STOI, PESQ, and SISDR scores. The score for each condition represents the average objective intelligibility or quality score of 1200 and 800 speech samples for three seen ($400\times3$) and two unseen ($400\times2$) noises, respectively. In general, the objective scores for enhanced speech are higher than unprocessed speech. Notably, the relative improvement is more pronounced at lower SNRs compared to higher ones. The results highlight that our proposed algorithm consistently outperforms baseline networks across every test condition. 

The CFTNet shows superior performance when compared to the unprocessed and DCCRN networks. A notable enhancement is observed when the bottom layer of the CFTNet is substituted with the DPRNN module, resulting in a competitive performance, especially at very low SNRs. The introduction of the DAT-RNN module into the CFTNet, culminating in the DAT-CFTNet network, brings about significant performance gains, particularly in the PESQ and SISDR metrics. Specifically, DAT-CFTNet reports improvements of +22.8\%, +113.4\%, and +10.62 dB in STOI, PESQ, and SISDR scores, respectively, over the unprocessed signal. Further advancements are seen with the DAT-CFTNet-F network, which demonstrates a marked improvement in quality over DAT-CFTNet. The result indicates that DAT-CFTNet-F achieves a relative improvement of +34.3\% and +6\% in the SISDR score over the DCCRN and CFTNet, respectively. Additionally, a +5.63\% boost in the PESQ score is observed when compared to the baseline networks. These findings highlight the effectiveness of the proposed DAT-CFTNet algorithm in enhancing speech intelligibility, and quality, and reducing speech distortion, especially for CI listeners. While the streamlined DAT-CFTNet-L model sees a dip in performance, it still outperforms the baseline DCCRN model and is competitive with the CFTNet model, marking it as a viable choice for use in CI systems for CI listeners.

\begin{table}
  \centering
  \caption{Ablation study of the proposed network. Average improvement across all noise types and SNRs is presented in terms of STOI, PESQ, SISDR, SOPM, IS, and LSD.}
 \includegraphics[width=0.8\linewidth]{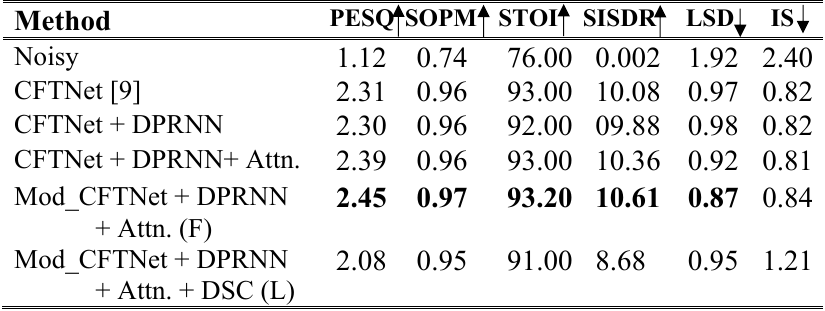}
 \label{table: Ablation_Score}
 \vspace{-15 pt}
 \end{table}
To elucidate the contributions of each component, a series of training and testing iterations were executed for the proposed model, with the resultant mean objective scores presented in Table~\ref{table: Ablation_Score}. Incorporating DPRNN yielded enhancements in both speech quality and distortion metrics, evident through elevated PESQ, SISDR, and LSD scores, while preserving consistent speech intelligibility. The DAT module within DPRNN demonstrated a notable +5.26\% relative improvement in SISDR and +4.85\% when compared to CFTNet. Subsequent enhancements were observed upon refining the proposed DAT-CFTNet by reducing the number of FTB in the encoder layer. Specifically, DAT-CFTNet-F exhibited relative improvements of +6.01\%, +5.26\%, and -10.31\% in PESQ, SISDR, and LSD scores over CFTNet, respectively. Furthermore, the streamlined architecture, DAT-CFTNet-L, significantly trimmed network parameters without a substantial decline in network performance.

\begin{figure}[t]
  \centering
  \includegraphics[width=0.45\textwidth]{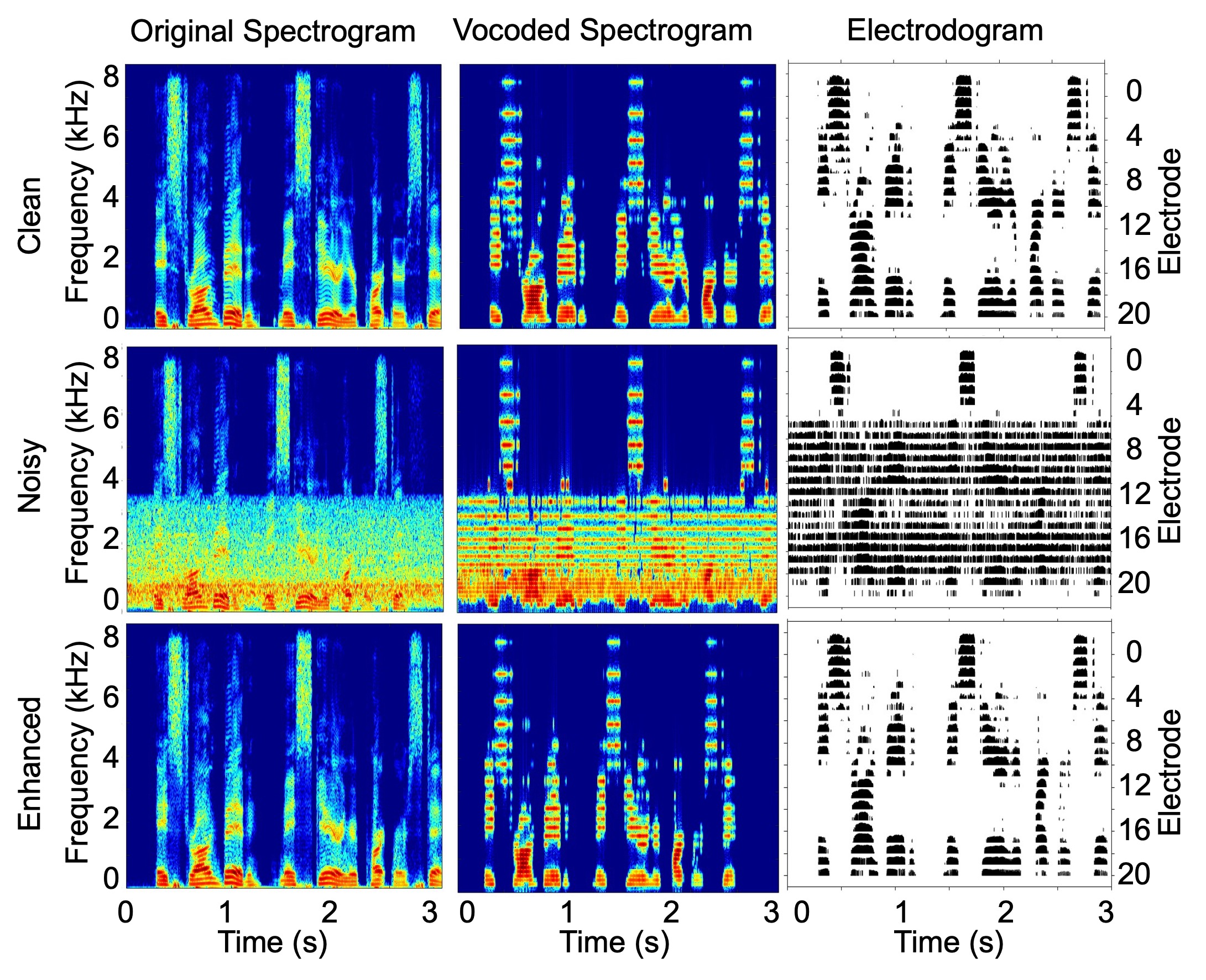}
  \caption{CI electrode stimulation response (electrodogram) with spectrograms with original and vocoded speech in car noise at SNR 0 dB for clean, noisy, DAT-CFTNet-F processed speech.}
  \label{fig: Spectrogram}
\vspace{-20pt}
\end{figure}

The significance of background noise suppression on speech perception for CI users is considerable, and SE techniques can serve to alleviate this issue. To evaluate the efficacy of the proposed algorithm for CI users, Fig.~\ref{fig: Spectrogram} displays the spectrograms and electrodograms of the processed signals. The original clean signal is deliberately contaminated with car noise at an SNR of 0 dB. The proposed network is then applied to enhance this noisy signal. Following this, an Advanced Combined Encoder (ACE) signal-processing strategy~\cite{skinner2002speech} is employed to simulate the CI-received signal for RF pulse generation and to produce the corresponding CI electrodograms. A standard CI parameter setting, involving the generation of biphasic electric RF pulse stimuli with 22 electrodes, is utilized \cite{hansen2019cci, ghosh2021cci}. The results reveal that the proposed DAT-CFTNet network is highly effective in attenuating noise while maintaining the harmonic structure of speech in both the spectrograms and CI electrodograms. 
\vspace{-20 pt}

\section{Conclusion}
\vspace{-10 pt}
This research has introduced an enhanced version of CFTNet, termed DAT-CFTNet, specifically designed to augment speech perception in real-world environments for both NH and CI listeners. By integrating a DAT-RNN module into the bottleneck layer of a complex-valued frequency transformation network, the network able to achieve significant improvements in both speech intelligibility and quality. Objective evaluations substantiated the effectiveness of DAT-CFTNet, revealing a notable +34.3\% increase in speech intelligibility and a +6\% enhancement in speech quality when compared to the baseline DCCRN network. In addition to the primary model, we also proposed a lightweight model, DAT-CFTNet-L, that reduces the model parameters by a factor of three. This variant maintains a balance between performance and computational efficiency, making it particularly suitable for resource-constrained applications. 
\vspace{-15 pt}

\section{Acknowledgment}
\vspace{-10 pt}
This work was supported by Grant No. R01 DC016839-02 from the National Institute on Deafness and Other Communication Disorders (NIDCD), National Institutes of Health.



\bibliographystyle{IEEEbib}
\bibliography{Reference_NM}

\end{document}